\documentclass[12pt,preprint]{aastex}

\newcommand\simlt{\lower.5ex\hbox{$\; \buildrel < \over \sim \;$}}
\newcommand\simgt{\lower.5ex\hbox{$\; \buildrel > \over \sim \;$}}

\slugcomment{Submitted to {\it Astrophysical Journal Letters},
DATE}

\begin{document}
\title{LIGO/VIRGO searches for gravitational radiation in hypernovae}
\author{Maurice H.P.M. van Putten}
\affil{MIT 2-378, Cambridge, MA 02139-4307}

\begin{abstract}
   A torus around a stellar mass Kerr black hole can emit about 10\% of the 
   spin-energy of the black hole in gravitational radiation, potentially
   associated with a gamma-ray burst. Wide tori may develop buckling modes by 
   the Papaloizou-Pringle instability and gravitational radiation-reaction 
   forces in the Burke-Thorne approximation. Gravitational wave experiments 
   may discover these emissions in a fraction of nearby supernovae. This 
   provides a test for Kerr black holes, and for GRB inner engines by comparison 
   with the de-redshifted durations of long GRBs.
\end{abstract}


\section{Introduction}

  Stellar mass black holes surrounded by a compact torus may represent catastrophic
  events such as core-collapse in a massive stars and black hole-neutron star
  coalescence. These scenarios have been considered as sources of 
  cosmological gamma-ray bursts \citep{woo93,pac91}. We may consider 
  black hole-torus systems and their emissions more generally, especially when
  the black hole is rapidly spinning. Their potential association with GRBs  
  provides observational constraints on their evolution.

  A torus around a rapidly rotating black hole may develop a state of suspended 
  accretion for the lifetime of rapid spin of the black hole \citep{mvp01a}. This
  points towards major energetic output in ``unseen" emissions, in gravitational 
  radiation, magnetic winds, thermal emissions  and neutrino emissions \citep{mvp02}. 
  The energy $E_{gw}$ emitted in gravitational radiation is expected to be about 
  $10\%E_{rot}$, i.e., 
  \begin{eqnarray}
  E_{gw} \simeq 6\times 10^{53}\mbox{erg}
  \label{EQN_E}
  \end{eqnarray}
  for a $10M_\odot$ black hole. This output (\ref{EQN_E}) 
  may be detected by the upcoming Laser Interferometric Gravitational Wave 
  Observatory LIGO \citep{abr92} and the French-Italian counter part VIRGO \citep{bra92}, 
  possibly in combination with any of the bar 
  or sphere detectors currently being developed. This provides 
  a calorimetric compactness test for Kerr black holes 
  \citep{mvp01}, and a means of identifying the inner engine of GRBs by comparison
  with de-redshifted durations of long GRBs.

  In the Woosley-Paczynski-Brown scenario 
  of hypernovae \citep{woo93,pac98,bro00,bro02}, core-collapse in rotating
  massive stars forms a Kerr black hole surrounded by a compact disk or torus. 
  Long GRBs correlate with star-forming regions \citep{blo00} and, hence, 
  young massive stars, possibly in binaries. 
  The rotating black hole may produce wide-angle ejecta back into the 
  interstellar medium leaving behind a soft X-ray transient 
  with a chemically enhanced companion star \citep{bro00}, 
  such as GRO J1655-40 \citep{isr99} 
  and V4641Sgr \citep{oro01}. This potential supernova and SXT association is
  important in identifying progenitors to GRBs and their inner engines.
  The beamed output of true GRB energies of $E_\gamma=10^{50-51}$ergs \citep{fra01}
  represents a minor energetic output for a long GRB from a rotating black hole.
  The potential for long gamma-ray bursts from rotating black holes suggests
  that GRB associated supernovae may emit bursts of gravitational radiation, e.g.,
  GRB 980425/1998bw \citep{gal98}, GRB 011121 \citep{blo02} and 
  GRB 011211 \citep{ree02}. 

  In this {\em Letter}, we suggest LIGO/VIRGO searches for bursts of gravitational 
  radiation from back hole-torus systems using upcoming continuous all-sky 
  supernovae surveys. Focusing on supernovae may serve to reduce data analysis
  by their well-determined coordinates, which includes distances.  
  The expected gravitational wave-spectrum is here identified with multipole
  moments in a wide torus due to a Papaloizou-Pringle instability, by extension of
  the theory for slender tori \citep{pap84,gol86} and including the secular effect 
  of gravitational radiation backreaction-forces in the Burke-Thorne 
  approximation \citep{tho69}.
 
\begin{figure}
\plotone{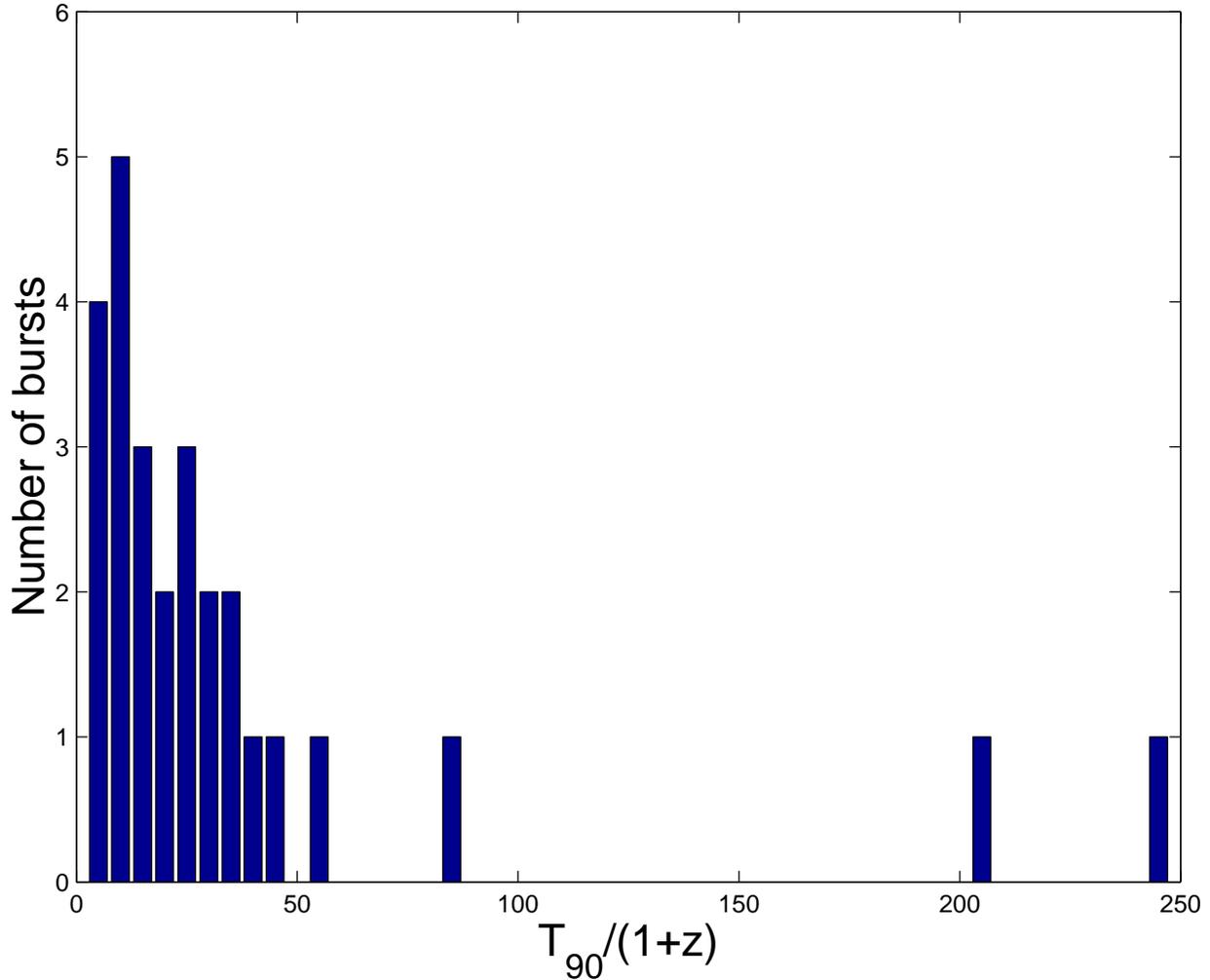}
\caption{Shown is the histogram of redshift corrected durations
of 27 long bursts with individually determined redshifts from their 
afterglow emissions (sample from Djorgovski et al., 2001, 
astro-ph/0107535, and references therein, and updated
(Djorgovski, 2002; Hurley, 2002).
The mean value of the durations of all bursts 
is 38s; the average is 23s without the long bursts GRB 980703 
$(T_{90}/1+z=400$s) and GRB 000911 $(T_{90}/1+z=243$s), which 
consist of two (in BATSE; one in Ulysses), respectively, three 
well-separated sub-bursts.
GRBs from rotating black holes are expected to be accompanied by
``unseen" emissions in gravitational radiation from the torus.
This predicts a similar distribution of durations for their bursts
of gravitational radiation with an expectation value of about one-half 
minute.}
\end{figure}

\section{GRB/GWBs from rotating black holes}
 
  A rapidly rotating black hole surrounded by a magnetized torus may
  develop a state of suspended accretion \citep{mvp01a}. An equilibrium
  magnetic moment in its lowest energy introduces 
  a high incidence of the black hole luminosity into the inner face of the torus 
  through a surrounding magnetosphere 
  with slip/no-slip boundary conditions, and enables support to an 
  open flux tube to infinity with slip/slip boundary conditions \citep{mvp02}. 
  
  Differential frame-dragging on an open flux tube may create a minor energy output 
  $E_{j}$ in low-sigma baryon-poor jets. In the proposed association to GRBs, 
  $E_j=E_\gamma/\epsilon$ represents a few permille of the rotational energy of the black 
  hole, where $\epsilon\sim 0.15$ denotes a fiducial efficiency of the conversion of
  kinetic energy-to-gamma rays. The associated horizon half-opening angle 
  $\theta_H\simeq 35^o$ provides a bound for the observed half-opening angle 
  $\theta_j$ on the celestial sphere, following collimation.
  A major energy output is expected to derive from the torus, upon catalyzing 
  spin-energy from the black hole into gravitational radiation (\ref{EQN_E}), winds, 
  thermal and MeV neutrino emissions. 
  We next consider the formation of a multipole mass moments in the torus, which
  defines its spectrum of gravitational radiation.

  \section{Multipole moments in wide tori}

  The effect of shear on the stability of a three-dimensional torus of incompressible fluid 
  can be studied about an unperturbed angular velocity $\Omega=\Omega_a(a/r)^q$,
  where $q\ge 3/2$ denotes the rotation index. In the inviscid limit, we 
  may consider, by Kelvin's theorem, irrotational modes in response to irrotational initial
  perturbations to the underlying flow (vortical if $q\ne2$). 
  We expand their harmonic velocity potential 
  \begin{eqnarray}
  \phi=\Sigma_n a_n(r,\theta,t) z^n, ~~\Delta\phi=0,
  \end{eqnarray}
  in cylindrical coordinates ($r,\theta,z$). The equations of motion can be conveniently 
  expressed in a local cartesian frame $(x,y,z)$ with angular velocity 
  $\Omega_a=M^{1/2}a^{-3/2}$ at $r=a$ about a central mass $M$, 
  where $x=r-a$, $\partial_x=\partial_r$ and $\partial_y=r^{-1}\partial_\theta$.
  (We can readily switch between cylindrical and cartesian coordinates in
  coordinate invariant expressions.) 
  Infinitesimal perturbations $\propto e^{im\theta-i\omega^\prime t}$ of frequency 
  $\omega^\prime$ as seen in the corotating frame at $r=a$ satisfy the linearized equations 
  of momentum balance. For an azimuthal quantum number $m$ and on the equatorial plane
  $z=0$, they are, in the notation of \citep{gol86},
  \begin{eqnarray}
  \begin{array}{rl}
  -i\sigma u -2\Omega v & = -\partial_r (h+\Phi),\\
  -i\sigma v + 2B u & = -ik (h+\Phi),
  \end{array}
  \label{EQN_EOM}
  \end{eqnarray}
  where $h$ denotes a perturbation of the unperturbed enthalpy
  $\partial_rh^e=\Omega^2 r -M/r^2$ about $z=0$, $\Phi$ a potential,
  $2B=(2-q)\Omega$, $k=m/r$, and $\sigma=\omega^\prime-m\delta\Omega$
  ($\delta\Omega=\Omega-\Omega_a$) 
  produces the Lagrangian derivative $D_t=-i\sigma + u\partial_x$. The equation
  of motion for the vertical velocity component, $-i\sigma w = -\partial_z h$,
  is decoupled in $z=0$. 
  In earlier linearized treatments \citep{gol86}, variations of $2B$ across the torus
  are neglected, limiting the discussion to narrow tori defined by $h^e(x_\pm)=0$.
  For wide tori, we here include $(2B)_x=-(q/r)2B$. The equations of motion 
  (\ref{EQN_EOM}) obtain
  \begin{eqnarray}
  (\partial_r^2+r^{-1}\partial_r-m^2r^{-2})a_0=qr^{-1}a_0^\prime
  \end{eqnarray}
  for an azimuthal mode number $m$. 
  Solutions symmetric about the equatorial plane satisfy
  \begin{eqnarray}
  \phi=a_0-\frac{z^2q}{2r}\partial_ra_0+O(z^4),~~a_0=r^{p_+}+ \lambda r^{{p_-}},
  \end{eqnarray}
  where $p_\pm=q/2\pm\sqrt{q^2/4+m^2}$ and $\lambda$ is a constant.

  The unperturbed inner and outer boundaries $x=x_\pm$ of the torus 
  are defined by the vanishing of the specific enthalpy: $h_0(x_\pm)=0$.
  In the limit of a narrow torus, this reduces to
  $h^e=({2q-3})\Omega_a^2 (b^2-x^2)/2$ ($-b\le x \le b$) \citep{gol86}.
  The Lagrangean condition $h(x_\pm)=0$ also in the
  perturbed state reduces to the
  two-point boundary condition (in the equatorial plane) \citep{gol86}
  \begin{eqnarray}
  0=D_t h =  -i\sigma h + u h_{x}^e.
  \end{eqnarray}
  The second equation in (\ref{EQN_EOM}) gives
  $h=i\sigma \phi -\Phi + ik^{-1}2B \phi_x,$
  so that \citep{gol86}
  \begin{eqnarray}
  k(\sigma^2\phi + i\sigma \Phi) + (2B \sigma  + k h_{x}^e)\phi_x = 0.
  \end{eqnarray}
  
  In the absence of a potential $\Phi$, the stability of the torus is described 
  in terms of a critical rotation index for each azimuthal quantum number $m$. 
  The boundary conditions $h^e=0$ become
  \begin{eqnarray}
  \epsilon\sigma^2 + 2B\sigma + k h^e_{x} = 0,
  \label{EQN_A1}
  \end{eqnarray}
  where $\epsilon={k\phi}/{\phi_x}$. The linearized equation obtained in the limit
  of small $\sigma$ in (\ref{EQN_A1}) corresponds to the slender torus approximation
  ($b<<a$) of Papaloizou-Pringle, or shallow water wave limit ($kb<<1$). About 
  $\omega=0$, this obtains a critical rotation index
  $q=\sqrt{3}$ for all $m$ \citep{pap84}. 
  Fig. 2 shows a complete stability diagram of critical values $q_c(b/a)$ for
  for azimuthal quantum numbers $m=1,2,3..$ and $0<b/a<1$, obtained by
  eliminating $\lambda$ in (\ref{EQN_A1}) and solving 
  for $q$ associated with double zeros of $\omega$ by continuation methods
  (e.g., \cite{keller}).

  The presented perturbations are buckling modes, 
  associated with the same sign of the radial
  velocity at the inner and the outer face. In contrast, 
  two-dimensional incompressible, vortical modes are defined by 
  $\sigma\Delta\psi=k(2B)_x\psi$ in terms of the streamfunction $\psi$ 
  ($u=\psi_y$ and $v=\psi_x$). They are
  generally singular with divergent azimuthal velocities when $\psi\ne0$ at the 
  turning point $\sigma=0$, though of finite net azimuthal momentum ($\psi$ remains
  continuous). These modes fall outside 
  the scope of the present discussion.
  \begin{figure}
  \plotone{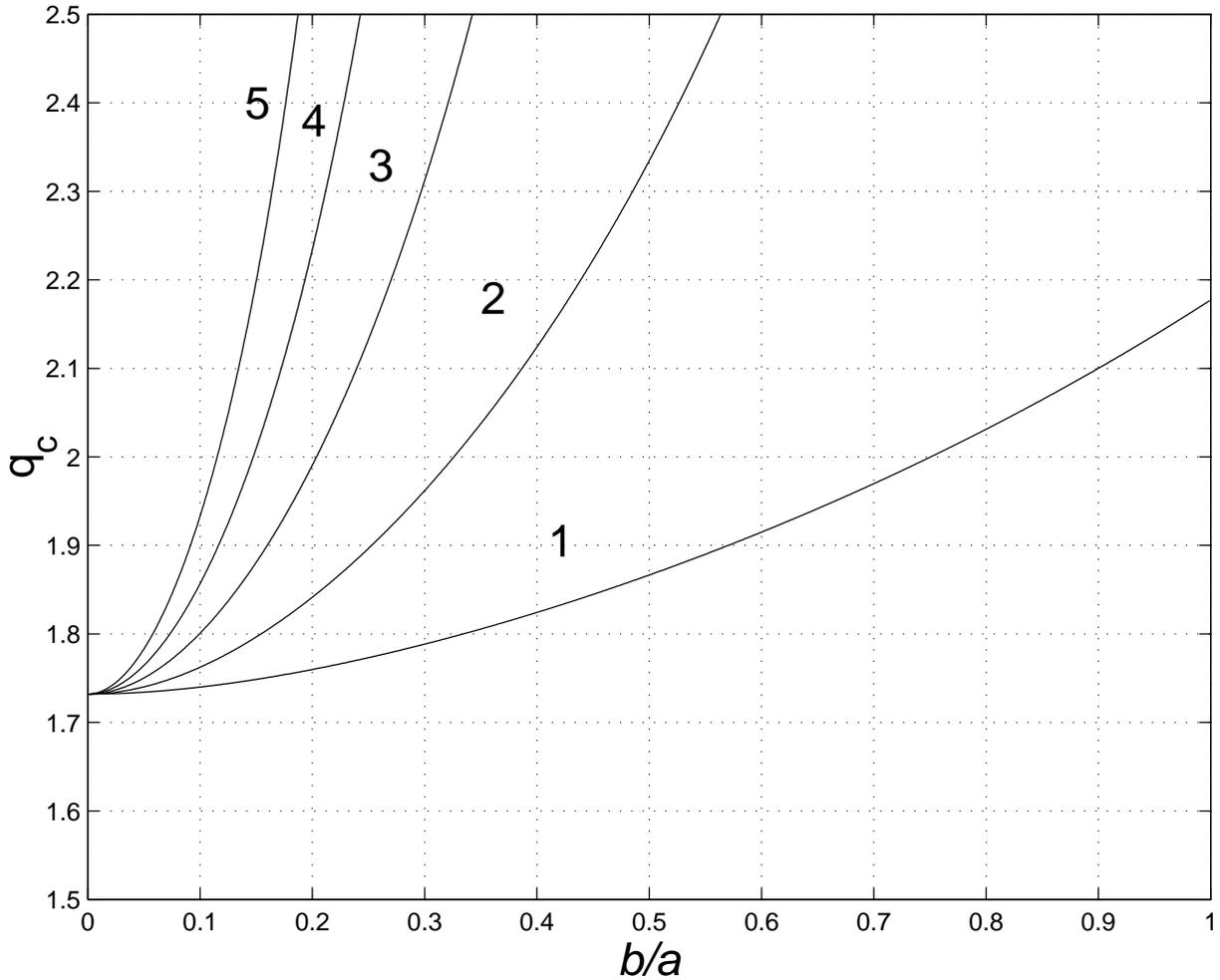}
  \caption{Diagram showing the neutral stability curves for buckling modes in a 
  torus of incompressible fluid, as an extension of the Papaloizou-Pringle 
  instability to large ratios of minor-to-major radius $b/a$. 
  Curves of critical rotation index $q_c$ are labeled with azimuthal quantum 
  numbers $m=1,2,..$, where instability sets in above and stability sets in below.
  Of particular interest is the range $q\le2$, where the $m=0$ mode is Rayleigh 
  stable. For $q=2$, the torus is unstable for $b/a$ less than
  $0.7385$ ($m=1$), $0.3225$ ($m=2$) and, asymptotically,  
  for $b/a\simeq 0.56/m$ $(m\ge3)$}
  \end{figure}  
  \begin{figure}
  \plotone{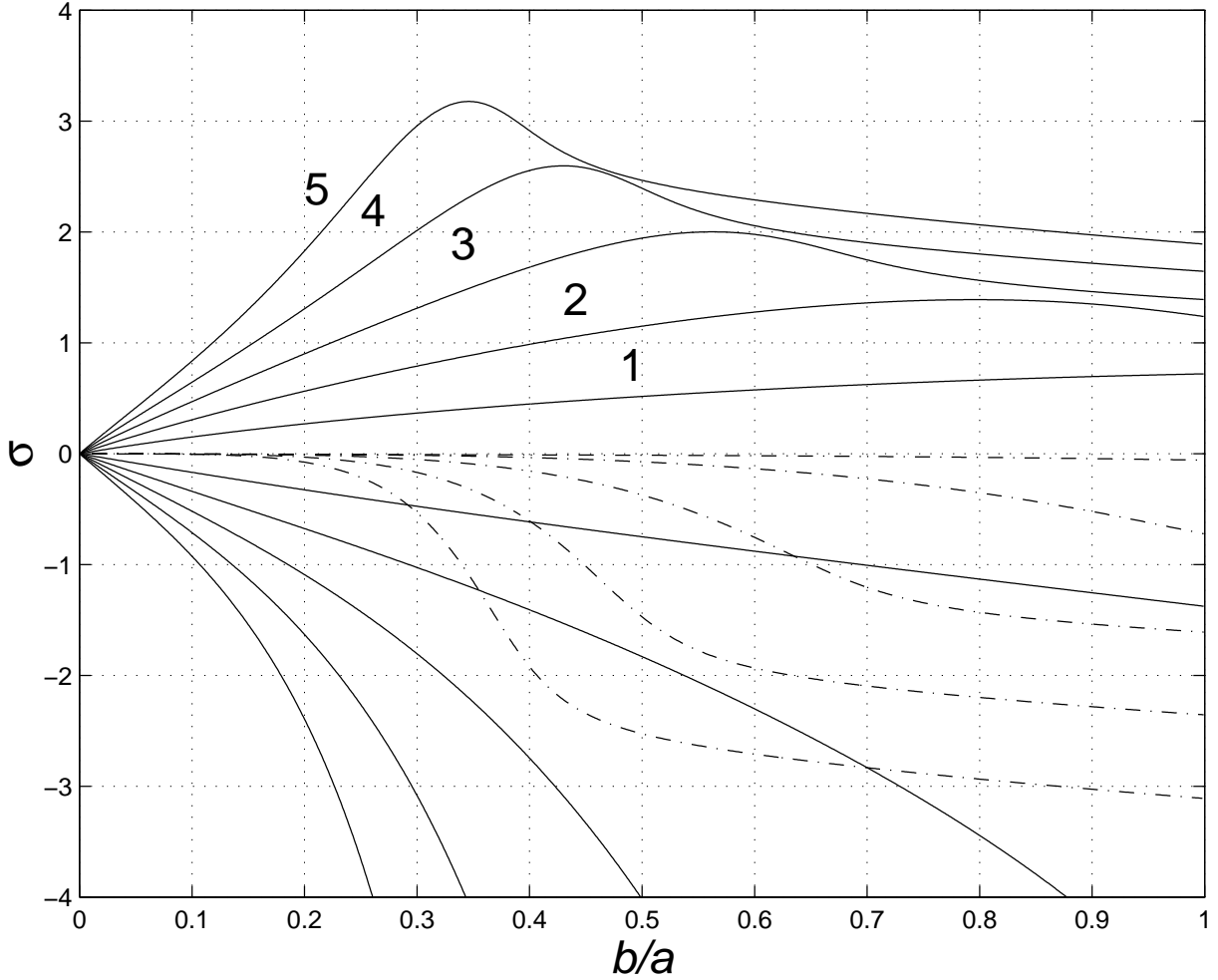}
  \caption{The frequency diagram of the pair of waves in a buckling mode on the
  neutral stability curves of a torus of incompressible fluid.
  The waves on the outer face are prograde (positive $\sigma$ curves; labeled
  for each azimuthal quantum number $m=1,2,\cdots,5$),  
  whereas the waves on the inner face are retrograde (negative $\sigma$).
  The dot-dashed lines refer to the frequency $\omega^\prime$ as seen in the corotating
  frame with angular velocity $\Omega_a=M^{1/2}/a^{3/2}$ of the torus at major
  radius $r=a$, where the highest (lowest) curve refers to $m=1$ ($m=5$). Note that
  up to $b/a\simeq 0.3$, $\omega^\prime$ remains close to zero. Hence, the observed
  frequency of the gravitational radiation as seen at infinity is close to 
  $m\Omega_a$ for low $m$.}
  \end{figure}  

  A quadrupole buckling mode emits gravitational radiation at twice the frequency
  $\omega=2(\Omega_a+\omega^\prime)\simeq 2\Omega_a$ (see Fig. 3). 
  It develops out of an internal flow
  of energy and angular momentum from the inner to the outer face of the torus,
  in which total energy and angular momentum of the wave remain zero. 
  The emitted gravitational radiation is therefore not extracted from the kinetic 
  energy in this pair of waves. This contrasts with radiation extracted  
  from single waves of frequency 
  $0<\omega<m\Omega_T$ by the Chandrasekhar-Friedman-Schutz instability. 
  It may be noted that this is equivalent to a positive entropy condition $\delta S>0$ 
  in the first law of thermodynamics $-\delta E=\Omega_T(-\delta J)+T\delta S$ for a 
  torus at temperature $T$, upon radiation of waves with specific angular momentum
  $\delta J/\delta E=m/\omega$ to infinity. See also \cite{sch80} on the entropy
  condition in the Sommerfeld radiation condition. 
  We proceed with the effect of gravitational radiation-reaction forces as follows. 

  \section{Gravitational radiation-reaction force}

  The backreaction of gravitational radiation consists of dynamical self-interactions 
  and radiation-reaction forces \citep{tho69,cha70,sch80}.
  For slow motion sources with weak internal gravity, the latter can be modeled by the 
  Burke-Thorne potential as it arises in the $2\frac{1}{2}$ post-Newtonian approximation
  \begin{eqnarray}
  \Phi_{BT}=\frac{1}{5}x_jx_k\left(I^{jk}-\frac{1}{3}I\delta^{jk}\right),
  \label{EQN_BT}
  \end{eqnarray}
  where $I_{jk}=\int \Sigma x_jx_k dxdy$ denotes the second moment tensor of
  the matter surface density $\Sigma$. This intermediate order does not introduce
  a change in the continuity equation (such as in the 2-nd order post-Newtonian
  approximation \citep{cha70,sch80,sch83}). In
  cylindrical coordinates $(r,\phi)$ ($x=r\cos\phi$ and $y=r\sin\phi$), 
  and for harmonic perturbations
  $\eta=\eta e^{2i\theta-i\omega t}=\eta e^{2i\theta}e^{-i\omega t}$ of the
  wave-amplitude, we have,
  in the approximation of a constant surface density $\Sigma$,
  \begin{eqnarray}
  I_{x_ix_j}  
   =  \Sigma\int_0^{2\pi} \int_{a+x_-+\eta_-}^{a+x_++\eta_+} x_ix_j dxdy. 
  \label{EQN_I}
  \end{eqnarray}
  Explicitly, $I_{xx}=({\pi\Sigma}/{2})
  \left[(a+x_+)^3\eta_+-(a+x_-)^3\eta_-\right],$
  which determines
  $\Phi_{BT}=(x^2I_{xx}+y^2I_{yy}+2xyI_{xy})/5=z^2I_{xx}/5,$
  where $z=x+iy$. Here, $I_{xx}=I_{xx}e^{-i\omega t}$ and $z^2$ comprises
  $e^{2i\theta}$ -- combined, $\Phi=\Phi e^{2i\theta-i\omega t}$. The harmonic
  time-dependence $e^{-i\omega t}$ derives from the integral boundaries
  in (\ref{EQN_I}), and hence applies to all components of the moment of inertia
  tensor. The linearized
  radiation-reaction force derives from the fifth time-derivative, i.e., 
  $\Phi=-i\omega^5\Phi_{BT}$ in the stability analysis of the previous section, 
  supplemented with the kinematic surface conditions $-i\sigma \eta = \phi_{x}$
  on the inner and outer boundaries $x=x_\pm$. Explicitly, we have
  $i\sigma_\pm \Phi(x_\pm) = i\beta (1+x_\pm)^2K(x_-,x_+),$
  where $K(x_-,x_+)=(1+x_+)^3 \phi_{x}(x_+)-(1+x_-)^3\phi_x(x_-)$
  and $\beta={\pi a\Sigma (\omega a)^5}/{10}$.
  A fiducial value $\beta\sim 10^{-4}$ obtains for
  for a torus mass $m_T=0.1M$ and a radius $a=3M$. Figure 4 shows the destabilizing
  effect of $\beta=2\times 10^{-4}$.
  \begin{figure}
  \plotone{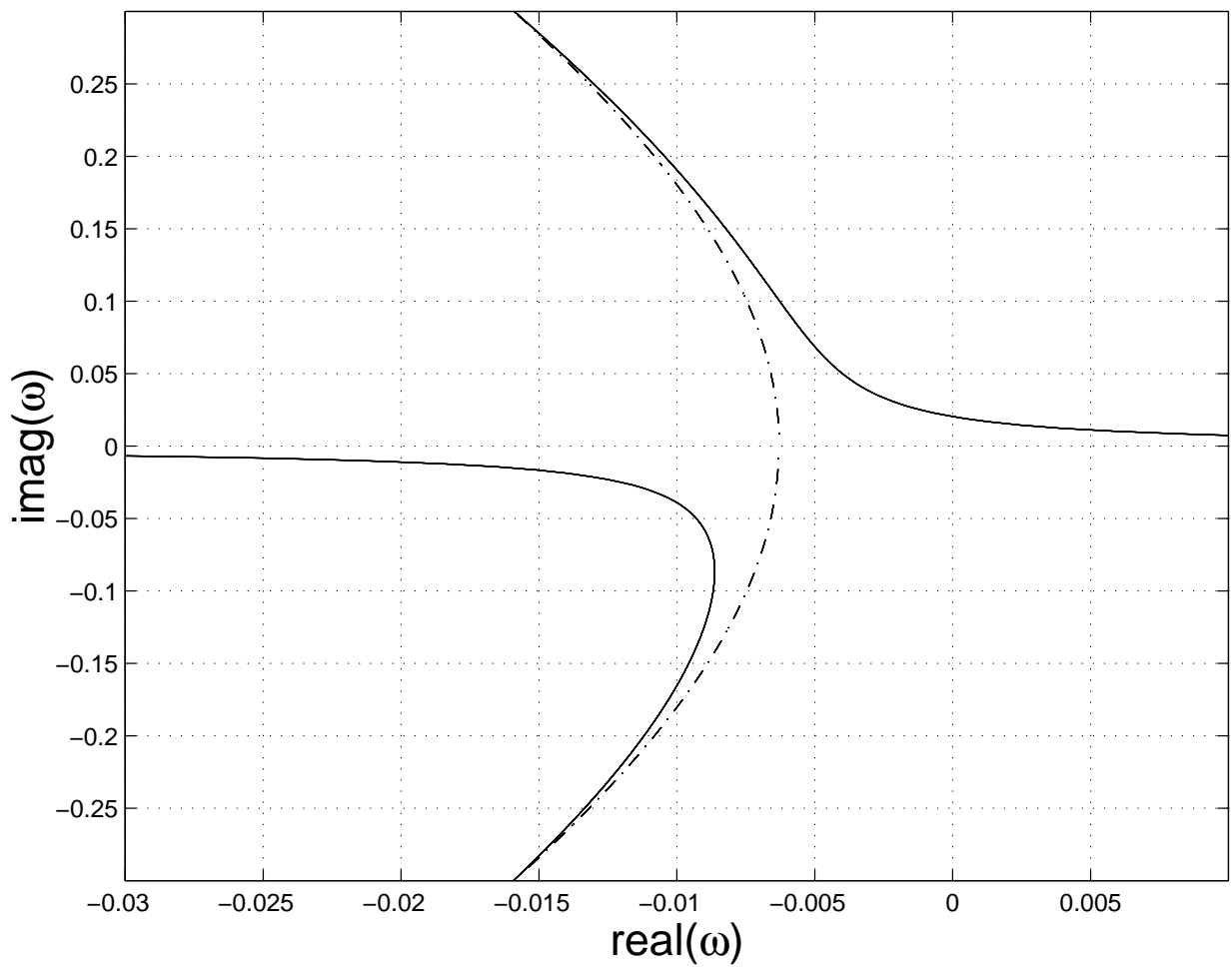}
  \caption{The complex frequency diagram of the frequency $\omega^\prime$ of the
  quadrupole moment in the torus in response to the radiation-reaction force. 
  The results are shown for a canonical value of
  $b/a=0.2$ and $\beta=2\times 10^{-4}$, corresponding to a torus mass
  of about one percent of the black hole mass. The dot-dashed curves
  are the asymptotes for $\beta=0$. This shows that gravitational
  radiation backreaction-forces renders contribute to an unstable quadrupole
  Papaloizou-Pringle buckling mode.}
  \end{figure}  
  
  The luminosity in quadrupole gravitational radiation,
  ${\cal L}_{gw}\simeq ({32}/{5})(\Omega_T M)^{11/3}(\delta m_T/M)^2$,
  results from a mass inhomogeneity $\delta m_T/M$ of a few permille \citep{mvp01}.
  The Papaloizou-Pringle instability may form a quadrupole shock
  on the inner and other face of the torus. Shocks amplitudes $\eta_\pm$ will 
  saturate by dissipation $\propto \eta_\pm^3$. Balance with energy input in the 
  suspended accretion state, being on the order of the luminosity in gravitational 
  radiation $\propto \eta^2_\pm$, leaves $\eta_\pm$ self-consistently to be on 
  the order of $\delta m_T/m_T$ of a few permille as well for canonical values 
  $m_T=0.1M_\odot$ and $M=10M_\odot$. 

  \section{Estimate of index of rotation}

  The index $q$ is related to the pressure $P$ in the torus. For a three-dimensional
  density $\rho$, the symmetric case in forementioned slender torus limit 
  $h=P/\rho=(2q-3)\Omega_a^2(b^2-x^2)/2$ gives
  \begin{eqnarray}
  q\simeq 1.5  + 0.75\times \left(\frac{a}{5M}\right)^3
  \left(\frac{M}{b}\right)\left(\frac{kT}{\mbox{5MeV}}\right),
  \label{EQN_Q} 
  \end{eqnarray}
  where $kT=$5MeV denotes a typical temperature of the torus in suspended accretion. 
  Hence, $q$ deviates from 3/2 by order unity.
  Furthermore, $\eta_T\simeq{f_H^2}/({\alpha+f_w^2+f_H^2})$, where $f_H, f_w$ 
  denote fractions of magnetic flux 
  in the inner and outer torus magnetospheres and 
  $\alpha=2qz-2f_w^2$ \citep{mvp01c}. Here, $z$ represents 
  viscosity between the inner and outer face of the torus in terms of the square
  of the ratio of radial turbulent magnetic flux to net poloidal magnetic flux.
  Fig. 2 shows $z=0.56$, derived from the number of unstable wave modes 
  in the limit of a slender torus. With
  a symmetric flux distribution $f_H=f_w=1/2$, we have 
  ${\Omega_T}/{\Omega_H}\simeq {1}/{4q}$.
  At a critical Rayleigh value $q=2$, the torus emits quadrupole radiation at 
  600Hz for a $7M_\odot$ black hole.

\section{LIGO/VIRGO searches for bursts in hypernovae}

Black hole-torus systems may form a class of transient sources with emissions
in several channels, possibly in association with GRBs.
For long GRBs, this suggests an event rate of about 1/yr within a distance of 
100Mpc in association with supernovae.

Continuous all-sky supernovae surveys may detect all or most nearby hypernovae.
Their coordinates provide a guide to searches for the proposed
bursts of gravitational radiation, by LIGO/VIRGO, bar (e.g. \citep{pog02}) or 
sphere detectors. This approach reduces the geometrical search space,
and may be augmented by surveys for transient sources in other wave-lenghts, e.g.,
in the radio \citep{lev02}. 

Searches for gravitational waves may find enhanced positive detection probability 
associated with supernovae, as the sum of false positive $P(+|-)$ and true positive 
detections $P(+|+)$, and in excess of $P(+|-)$ away from supernova events.
If all long GRBs stem from hypernovae, we expect a local supernovae-to-hypernovae ratio 
of a few hundred. The excess $P(+|+)$ becomes apparent for a signal-to-noise ratio 
better than two. Alternatively, a priori selected
hypernova candidates from the parent supernova sample, e.g., by detection of late 
time radio afterglows, would give an apparent excess $P(+|+)$ at lower signal-to-noise 
ratios. Complemented with distance estimates to supernovae or hypernovae, this 
provides a test for Kerr black holes \citep{mvp01}. We may test for GRBs from rotating
black holes by comparing the durations of bursts of gravitational radiation with
with the de-redshifted durations of long GRBs.

{\bf Acknowledgement.} The author acknowledges stimulating discussions with
D. Frail, C.-H. Lee, R. Weiss and A. Levinson, and thanks G. Djorgovski and 
K. Hurley for kindly providing the GRB sample for Fig. 1. This research is 
supported by NASA Grant 5-7012 and an MIT C.E. Reed Fund.

\end{document}